# BLACKBOX TROJANISING OF DEEP LEARNING MODELS : USING NON-INTRUSIVE NETWORK STRUCTURE AND BINARY ALTERATIONS

JONATHAN PAN

Home Team Science and Technology Agency, Singapore
E-MAIL: Jonathan_Pan@htx.gov.sg

**Abstract:**

Recent advancements in Artificial Intelligence namely in Deep Learning has heightened its adoption in many applications. Some are playing important roles to the extent that we are heavily dependent on them for our livelihood. However, as with all technologies, there are vulnerabilities that malicious actors could exploit. A form of exploitation is to turn these technologies, intended for good, to become dual-purposed instruments to support deviant acts like malicious software trojans. As part of proactive defense, researchers are proactively identifying such vulnerabilities so that protective measures could be developed subsequently. This research explores a novel blackbox trojanising approach using a simple network structure modification to any deep learning image classification model that would transform a benign model into a deviant one with a simple manipulation of the weights to induce specific types of errors. Propositions to protect the occurrence of such simple exploits are discussed in this research. This research highlights the importance of providing sufficient safeguards to these models so that the intended good of AI innovation and adoption may be protected.

## 1. Introduction

Artificial Intelligence (AI) is gaining in adoption in many scientific and technology domains. This is especially so with the advancements made by Deep Learning (DL) algorithms within the AI domain. The World Economic Forum recognizes AI as one of the key technology pillars driving the world towards the Fourth Industrial Revolution [22]. Such technology deployments now assume vital roles in keeping our lives and society safe and secure with deployments in health [23] and security [24].

Both Artificial Intelligence and Deep Learning have also been touted to be the game changer in the asymmetric cyber security landscape with the opportunity for cyber defenders to level up their capabilities to deal with the technological advancement of adversarial cyber attackers [4]. However as with all technologies, there will be associated forms of vulnerabilities to which adversarial actors may exploit to cripple, defunct and even turn these technology tools into adversarial ones for their own malicious use [2].

Artificial Intelligence has reportedly been used as offensive tools to identify vulnerabilities in systems and automate cyber attacks [5]. Brundage et. al further explores the possible use of AI and DL not only to be an offensive tool but as a trojaned tool disguised as a benign tool to an unassuming victim but transforms into the malicious Malware to launch its attack when triggered. The key question is how does one turn a benign model into a trojan.

This preliminary research work studies a blackbox approach to trojanise a pre-trained model to induce malicious inferences with two simple steps. The first is a simple modification to the network structure to the target model. Our work demonstrates that this could be done without the need to alter the original model or have intrinsic knowledge of the model construct or network structure. The second step that relates to the trigger mechanism to turn a benign model to induce the intended model inference errors involves precise weights manipulation. This simple alteration of specific weight values, also known as binary level attack trigger [12], that takes $O(1)$ time to execute.

The next section of this paper provides background information about adversarial adoption of AI and DL. This is followed by related research work in the techniques used to trojanise deep learning models and protective techniques available to deal with such threats. The description of the new form of trojanising technique follows with a description of the validation tests and an analysis. The paper continues with a discussion of possible infection vectors and propositions for protective mechanisms to prevent this malicious exploit and concludes with a conclusion to this research work.

## 2. Adversarial Artificial Intelligence

Cyber security researchers have studied the use of Artificial Intelligence algorithms like Reinforcement Learning (RL) [1] and Generative Adversarial Networks (GAN) [2] to generate Malware that could evade detection from anti-malware scanners. DeepPhish by Bahnsen et. al [4] demonstrated that their Long Short-Term Memory networks (LSTM) could effectively evade Phishing URL detectors. At BlackHat 2018, IBM presented their advanced malware based on Deep Learning algorithms called DeepLocker [3] that could evade detection from most anti-malware detectors and trigger the delivery of its malicious payload only when this AI malware recognizes the targeted victim using its facial recognition capabilities. As it is based on Deep Neural Networks, the standard malware forensics techniques of using reverse engineering to study the Malware to build counter-measures is not possible due to the 'blackbox' characteristics of such networks. These research work and demonstrators reflect the enhanced adversarial capabilities with the use of Artificial Intelligence algorithms especially those with Deep Learning algorithms.

## 3. Related Work

### 3.1. Trojaned Deep Learning

Another form of Adversarial Artificial Intelligence would be trojanised the AI models. There are a number of leading research work in studying how Deep Learning models could be trojaned. One popularly mentioned approach is to train models on the onset with corrupted or poisoned training datasets [9],[10],[11]. Chen et. al [6] demonstrated that a targeted backdoor could be created in Deep Learning systems using such data poisoning strategy by injecting poisonous samples into the training set to achieve adversarial objectives without any knowledge of the model construct and training dataset with the trigger key that would be unnoticeable by humans. Liao et. al [8] proposed another novel approach to data poisoning by designing different forms of stealthy perturbation masks applied to the training or update datasets as backdoor applied before model training or during model updates. Such techniques of trojanising the target model affects the induction bias of the model.

Clements and Lao [3] proposed a technique of inserting malicious hardware trojan into the implementation of a neural network classifier to induce the desired perturbation or affect the activation output from custom hardware trigger and circuitry. This is a form of side-channel attack against the target model.

Zou et. al [7] proposed their novel approach to insert modification to the targeted pre-trained model at the neuron level. Their approach also known as PoTrojan involved the additional extra neuron or neurons and their corresponding connection synapses and weights that would induce some variations to the error rates of the targeted model. However, their research proposition required significant modification and targeted training of the affected layers to achieve the intended malicious outcomes from the trojaned models. Another similar work is by Guo et. al [14] that proposed alterations to the original benign network and required training to embed the hidden trojan horse model. Such techniques involve the alteration of the target model.

Finally, trojanised behaviours can be induced by altering the weights of the target models. This is also known as binary level attack [12]. Rakin et. al [13] proposed this form of neural networks based trojan attack by performing a bit-flip attack on certain vulnerable bits of the targeted Deep Neural Networks using a gradient ranking approach that induces all inputs towards one class. Our approach involves the combination of two approaches. The first is the model alteration while maintaining the integrity of the original network regardless whether the trojaned model is in either a benign or malicious mode. This alteration is only applied after the target model has completed its training phase. The second involves the trigger mechanism uses the binary level approach to switch between modes. Unlikely many of research discoveries, our research work demonstrates that a model can be trojaned without the need to disrupt inductive bias of the model through its ingested data but through the disruption of the model bias while preserving the structural integrity and inference accuracy of the original target model when it is either benign and malicious modes.

### 3.2. Protection against Trojan Deep Learning

Most research work done are focused on detecting and defending trojaned model induced by poisoned training or inference datasets. Gao et. al's [17] proposed the use of entropy measurement of entropy in the model's prediction against the provided inputs. Baluta et. al [15] developed a prototype tool called NPAQ that checks whether a set of neural networks (N) upholds a property (P). If a trojaned model exists, the measurement from NPAQ can be used to verify that the Trojan has been removed after the model is retrained on benign datasets. He et. al [16] uses the approach of measuring the sensitivity of the model when using a small set of 'sensitive-samples'. Liu et al. [20] suggested that using trained anomaly detection classifiers could detect neural trojans however with a high false positive rate.

These protective techniques are predominantly focused on detecting effects of poisoned data on the models under observation. While such techniques may be applied to our trojaned model when it is in malicious mode, false negative outcomes would be reported when it is applied to our model when it is in benign model.

## 4. Trojanising Deep Learning Models

In our research, we demonstrate how a model could be trojaned to produce malicious inference through two simple steps. The first involves extending the target model as part of the malicious payload design and the other involves binary manipulation of the weights in the extended part of model as part of the trigger mechanism. Our current research work was applied to target models based on popular Deep Learning models using Convolutional Neural Networks (CNN) that performs feature extraction and capped by Fully Connected layers of standard Neural Networks (FCN) for the classification tasks. These are a number of these frequently used classification models with pre-trained weights readily available.

4.1. Malicious Payload Design

The malicious payload design involves the addition of one layer on top of the target model. This extension is placed on top of the target model's output neurons. The input and output dimensions of this additional layer is the same as the dimensions of the output layer of the target model. Hence a pre-trained Imagenet model having 1000 outputs would have one additional layer of 1000 neural input and outputs which can be expressed as expressed in the following mathematical equations.

$$\widehat{y^j} = \sigma^j(x) \quad (1)$$

$$\widehat{y^t} = \sigma^t(\widehat{y^j} * W^t) \quad (2)$$
$$= \sigma^t(\sum_{i=0}^{n} \widehat{y_i^j} * w_i^t)$$

Hence the trojaned model takes the output $\hat{y}$ of the target model $\sigma^j$ as the inputs to produce the trojaned output $\widehat{y^t}$. The matrix $W^t$ represents the weights of the additional layer without bias. Equation 2 details the mathematical expression for the additional layer. A linear function is used for $\sigma^t$ hence when the appropriate matrix values are applied, the output of this layer has linear characteristics. The key design intent of our trojanisation approach to induce deterministic precise malicious outputs.

4.2. Malicious Payload Development

In this research demonstration, our trojaned model can operate in four modes. The first mode is benign mode that causes the trojaned model to behave normally. The prediction and accuracy outputs remain the same as the original target model would produce. In order to have this outcome, the additional layer uses an identity mapping $\sigma^t(\widehat{y^j}) = \widehat{y^j}$ with the weights $W^t$ set to be the identity matrix. The output of the original model is carried through without change to the output of additional trojaned layer as illustrated in Figure 1(a).

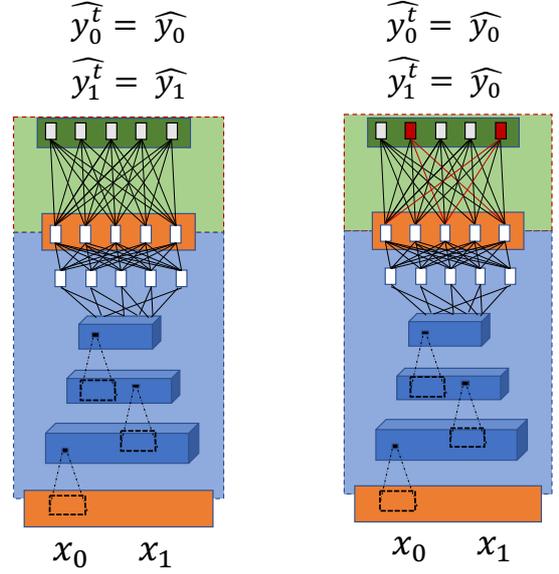

Figure 1. The left model (a) represents the Trojaned model in Benign Mode. The right model (b) represents the Trojaned model in Type I error.

The second mode is to induce of type I error for the targeted class hence resulting in False Positive outputs. The trojaned model will intentionally infer a positive identification for a specific invalid input. This is done by altering the weights' value to map the inputs (from the original model) to the intended outputs. Figure 1(b) illustrates the trojaned model's inference behavioral output.

The third mode is the induction of type II error that represents False Negatives as illustrated in Figure 2 (a). The trojaned model will classify the selected class as another class while validly inferring the other classes correctly. The final mode involves switching the classification of two selected classes Figure 2(b). Like first mode with type I error, both of these latter modes are configured through the manipulation of the identity matrix to map the outputs of the original model to the trojaned outputs.

As the payload design requires specific values of the matrix $W^t$ to be altered hence the change of modes may be done very quickly in *O(1)* time with no additional preparation, like retraining, is required. Additionally, the trigger to switch between modes could be done with a small payload of a few bytes dependent on the byte size of specific weight values.

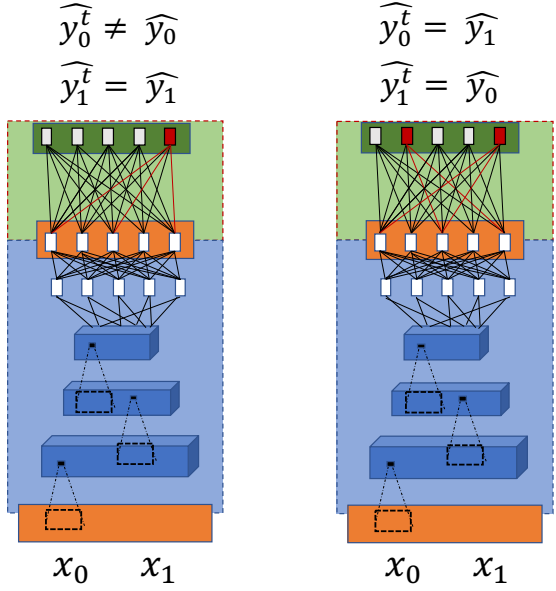

Figure 2. The left model (a) represents the Trojaned model in Type II error Mode. The right model (b) represents the Trojaned model in Switched Classification error.

### 4.3. Delivery of Trigger Mechanism

As mentioned earlier, the simple change of the weights of additional layer induces the change of state or mode of the trojaned model. This simple change of a few bytes has many possible delivery mechanisms options. The trigger could be delivered while the model is running as suggested by Rakin et al [13]. Alternatively, the change could be applied to a persistent form of the weights in the file system. In this research work, the focus is on the payload design hence a simple trigger mechanism of direct codification of the assigned weight values was used as part of the validation step.

## 5. Methodology and Analysis

### 5.1. Validation Test

In our research demonstration, we used three popular classification models with readily available trained weights on ImageNet dataset from open source repositories. They are VGG16, VGG19, InceptionV3 [18][19]. Our validation test involves the task of classifying ImageNet images. In order to demonstrate the research objective that the selected target Deep Learning Image Classification models can be trojaned, we define three steps that we would apply to the selected targeted models.

The first step is the preparation step that involves modifying the target models. This started with the instantiation of the target model and its ImageNet pre-trained weights from the Internet. We froze the weights of the target model hence rendering the model non-alterable throughout the validation test. We then applied the additional layer with unfrozen weights to the top layer of the target model. The weights of $W^t$ is initialized to the identity matrix. This sets the trojaned model to be benign mode.

The next step is to set the weights of the malicious additional layer to induce the desired behavior for the test namely benign, Type I error, Type II error, Switched Classification Error as described in the previous section. For the benign mode, there is no further step is needed as the trojaned model is initially configured with its identity matrix to behave per normal. For the other modes, the primary class is first selected $x^1$. This primary class represents the affected inferred class from the classification inference. Another class, that represents the secondary class $x^2$, is selected to assume the supporting role to the primary class. This secondary class facilitates the generation of type I error by having the trojaned model inferring the secondary class as the primary class. The choice of the secondary class is also needed for Switched Classification error mode so that the inferred identity of the primary class is swapped with the secondary class. For Type II error mode to induce False Negative inference will only require the primary class. The expected output from the trojaned models will be validated using the following formulations where $i = 1$ and $i = 2$ represents the primary and secondary classes respectively.

$$\widehat{y^t} = \{\sigma^j(x^i) \text{ when } i \neq 1 \text{ \& } i \neq 2 \quad (3)$$

$$\widehat{y^t} = \begin{cases} \sigma^j(x^i) \text{ when } i \neq 2 \\ \sigma^j(x^1) \text{ when } i = 2 \end{cases} \quad (4)$$

$$\widehat{y^t} = \begin{cases} \sigma^j(x^{i \neq 1}) \text{ when } i = 1 \\ \sigma^j(x^2) \text{ when } i = 2 \end{cases} \quad (5)$$

$$\widehat{y^t} = \begin{cases} \sigma^j(x^2) \text{ when } i = 1 \\ \sigma^j(x^1) \text{ when } i = 2 \end{cases} \quad (6)$$

The final step is to assess the classification outputs of the trojaned model in its various modes. We used the Tiny Imagenet [21] test dataset to assess the trojaned models. In our test, we selected three different pairs of primary and secondary classes and evaluated these models based on the number of True Positives, True Negatives, False Positives and False Negatives to facilitate consistent reporting of experimental results. Additionally, we evaluated the effects of the trojaned models with their other inference performance for non-affected classes to assess the premise that such trojanisation techniques has little effects of the rest of the trojaned models' inference performance.

As this evaluation is to compare the inference effects of the trojaned model over the original model, hence the evaluation is done with the ground truth taken from the output of the original model $\hat{y}$ and compared with the output from the trojaned model $\widehat{y^t}$.

5.2. Results and Analysis

The following were the observations gathered from the test. The accuracy measurements based on the equations [3] to [6].

| Test Case | Mode | Other Classes | Targeted Classes |
|---|---|---|---|
| 1 | 1 | 100.0% | 100.0% |
| 1 | 2 | 100.0% | 100.0% |
| 1 | 3 | 98.7% | 92.5% |
| 2 | 1 | 100.0% | 100.0% |
| 2 | 2 | 100.0% | 100.0% |
| 2 | 3 | 98.9% | 91.9% |
| 3 | 1 | 100.0% | 100.0% |
| 3 | 2 | 100.0% | 100.0% |
| 3 | 3 | 96.5% | 93.1% |

Table 1 – Trojaned VGG16 Model

| Test Case | Mode | Other Classes | Targeted Classes |
|---|---|---|---|
| 1 | 1 | 100.0% | 100.0% |
| 1 | 2 | 100.0% | 100.0% |
| 1 | 3 | 99.0% | 93.6% |
| 2 | 1 | 100.0% | 100.0% |
| 2 | 2 | 100.0% | 100.0% |
| 2 | 3 | 98.8% | 94.8% |
| 3 | 1 | 100.0% | 100.0% |
| 3 | 2 | 100.0% | 100.0% |
| 3 | 3 | 97.7% | 93.5% |

Table 2 – Trojaned VGG19 Model

| Test Case | Mode | Other Classes | Targeted Classes |
|---|---|---|---|
| 1 | 1 | 100.0% | 100.0% |
| 1 | 2 | 100.0% | 100.0% |
| 1 | 3 | 100.0% | 98.8% |
| 2 | 1 | 100.0% | 100.0% |
| 2 | 2 | 100.0% | 100.0% |
| 2 | 3 | 100.0% | 100.0% |
| 3 | 1 | 100.0% | 100.0% |
| 3 | 2 | 100.0% | 100.0% |
| 3 | 3 | 100.0% | 100.0% |

Table 3 – Trojaned InceptionV3 Model

The results observed was as expected with the trojaned models generating the corresponding errors. With the non-selected classes, the inference accuracies were maintained with good accuracy outcomes while the affected classes were wrongly induced as expected. The accuracies for both selected and other classes in Mode 3 fluctuated slightly. This may be due to the effects of altered model biases when two classes are involved.

6. **Infection and Protection**

6.1. Infection & Trigger Vectors

One possible infection vector to trojanise the targeted Deep Learning model is during the development of the model. The trojanisation of the target model may also happened when the model is already deployed in production however such modification of the network structure and its persistent weights would draw more attention to the induced change if the model and its weights are monitored closely for tampering.

The trigger vector to initiate the change of mode of the trojaned model from its current mode to another may be delivered in a number of ways given the small signature of change to be induced. Examples may include the use of customized Malware with precise weight alteration mechanism delivered to the model during execution or at rest with the persistent form of the weights or hyperparameters that exist in the form of files or database entries.

6.2. Protection from Model Trojanisation

This research demonstration highlights the need to protect targeted models from internal and external attacks. The first order of protection needed is to secure the integrity of the model structure from insidious modification. As described in the previous section about the infection and trigger vectors, protective safeguards over the integrity of model structure is needed at all time through its lifecycle that includes during development phase with its design, training and validation. The model will also need to protect from unauthorized modification during production phase with regular integrity checks on the model structure.

Additionally, in the event that the targeted model is still already trojaned but to deal the risk of trigger delivery, close monitoring and continual integrity checks of the weights or hyperparameters will need to be applied.

7. **Conclusion**

In this preliminary research, we demonstrated how a pre-trained target image classification model can be trojaned to induce specific errors using a simple modification to model. The trigger may be delivered through a lightweight model weights modification. Our approach induces the maliciously directed model biases at the binary level without the need to induce inductive bias from poisoned data. Additionally, unlike other forms of model trojanisation, our approach induces near deterministic malicious outcomes with its specific weight value manipulation. This research demonstration highlights the need to protect AI models and reap their beneficial gains. This could be done by applying tight change control protection to ensure that network structure of the model and its weights are protected at all stages of the lifetime of model's existence from both insider and external threats. Regularly integrity checks are needed to protect the models from unauthorized tampering or alterations.

**References**

[1] Anderson, H. S., Kharkar, A., Filar, B., Evans, D., and Roth, P. 2018. Learning to evade static pe machine learning malware models via reinforcement learning. arXiv preprint, arXiv:1801.08917.

[2] W. Hu and Y. Tan. Generating adversarial malware examples for black-box attacks based on GAN. arXiv preprint arXiv:1702.05983, 2017.

[3] D. Kirate, J. Jang and M. P. Stoecklin, "DeepLocker – Concealing Targeted Attacks with AI Locksmithing", BlackHat 2018, https://www.blackhat.com/us-18/briefings/schedule/index.html#deeplocker---concealing-targeted-attacks-with-ai-locksmithing-11549.

[4] Bahnsen, A.C.; Torroledo, I.; Camacho, L.D.; Villegas, S. DeepPhish: Simulating Malicious AI. In Proceedings of the Symposium on Electronic Crime Research, San Diego, CA, USA, 15–17 May 2018.

[5] M. Brundage et al., "The malicious use of artificial intelligence: Forecasting, prevention and mitigation", arXiv:1802.07228, Feb 2018.

[6] X. Chen, C. Liu, B. Li, K. Lu, and D. Song, "Targeted backdoor attacks on deep learning systems using data poisoning", ArXiv e-prints, abs/1712.05526, 2017.

[7] M. Zou, Y. Shi, C. Wang, F. Li, W. Song and Y. Wang, "Potrojan: powerful neural level trojan designs in deep learning models", ArXiv preprint arXiv:1802.03043, (2018)

[8] C. Liao, H. Zhong, A. Squicciarini, S. Zhu, and D.J. Miller, "Backdoor embedding in convolutional neural network models via invisible perturbation", ArXiv, arxiv:1808.10307, Aug. 2018.

[9] L. Munoz-Gonzalez, B. Biggio, A. Demontis, A. Paudice, V. Wongrassamee, E. C. Lupu and F. Roli, "Towards poisoning of deep learning algorithms with back-gradient optimization", Procedings of the 10th ACM Workshop on Artificial Intelligence and Security, ACM, 2017, pp. 27-38.

[10] C. Yang, Q. Wu, H. Li and Y. Chen, "Generative poisoning attack method against neural networks", ArXiv, preprint arXiv:1703.01340, 2017.

[11] M. Barni, K. Kallas and B. Tondi, "A new backdoor attack in CNNs by training set corruption without label poisoning", ArXiv, preprint arXiv:1902.11237, 2019.

[12] Y. Liu, A. Mondal, A. Chakraborty, M. Zuzak, N. Jacobsen, D. Xing and A. Srivastava, "A Survey on Neural Trojans", International Association for Cryptologic Research, https://eprint.iacr.org/2020/201.pdf, 2019.

[13] A.S. Rakin, Z. He, and D. Fan, "TBT: Targeted Neural Network Attack with Bit Trojan", ArXiv, arXiv preprint arXiv:1909.05193 (2019).

[14] C. Guo, R. Wu, and K.Q. Weinberger, "TrojanNet: Embedding Hidden Trojan Horse Models in Neural Networks", ArXiv, arXiv preprint arXiv:2002.10078 (2020).

[15] T. Baluta, S. Shen, S. Shinde, K.S. Meel and P. Saxena, "Quantitative Verification of Neural Networks And its Security Applications", ArXiv, preprint arXiv:1906.10395, 2019.

[16] Z. He, T. Zhang and R. Lee, "Sensitive-Sample Fingerprinting of Deep Neural Networks", Proceedings of the IEEE Conference on Computer Vision and Pattern Recognition, pp. 4729-4727, 2019.

[17] Y. Gao, C. Xu, D. Wang, S. Chen, D.C. Ranasinghe and S. Nepal, "STRIP: A Defence Against Trojan Attacks on Deep Neural Networks", ArXiv, preprint arXiv:1902.06531, 2019.

[18] K. Simonyan and A. Zisserman, "Very deep convolutional networks for large-scale image recognition", ICLR, 2015.

[19] C. Szegedy, V. Vanhoucke, S. Ioffe, J. Shlens, and Z. Wojna, "Rethinking the inception architecture for computer vision", CVPR, 2016.

[20] Y. Liu, Y. Xie, and A. Srivastava, "Neural trojans", IEEE International Conference on Computer Design (ICCD), IEEE, pp 45–48, 2017.

[21] J. Wu, Q. Zhang, G. Xu, Tiny imagenet challenge, cs231n, Stanford University.

[22] K. Schweb, "The Fourth Industrial Revolution: what it means, how to respond", World Economic Forum, 2016, https://www.weforum.org/agenda/2016/01/the-fourth-industrial-revolution-what-it-means-and-how-to-respond/.

[23] T. Davenport and R. Kalakota, "The potential for artificial intelligence in healthcare", Future Healthc. Journal, 2019, 6, 94–98.

[24] H. Chen, "AI and Security Informatics," IEEE Intelligent Systems, vol. 25, no. 5, 2010, pp. 82–90.